\documentclass[prl,twocolumn,showpacs,preprintnumbers,amsmath,amssymb,superscriptaddress]{revtex4}
\usepackage{amssymb}
\usepackage{color}

\def \Grenoble{Laboratoire National des Champs Magn\'etiques Intenses, CNRS-UJF-UPS-INSA, 38042 Grenoble, France}
\def \Prague{Institute of Physics, Faculty of Mathematics and Physics, 12116 Prague, Czech Republic}
\def \Warsaw{Institute of Physics, Polish Academy of Sciences, 02-668 Warsaw, Poland}

\usepackage{epsfig}
\usepackage{amsmath}

\begin{document}

\title{Enhancement of the spin-gap in fully occupied two-dimensional Landau levels}
\author{J. \surname{Kunc}}
\email{jan.kunc@grenoble.cnrs.fr}
\affiliation{\Grenoble}
\affiliation{\Prague}
\author{K. \surname{Kowalik}} \affiliation{\Grenoble}
\author{F. J. \surname{Teran}} \affiliation{\Grenoble}
\author{P. \surname{Plochocka}} \affiliation{\Grenoble}
\author{B. A. \surname{Piot}} \affiliation{\Grenoble}
\author{D. K. \surname{Maude}} \affiliation{\Grenoble}
\author{M. \surname{Potemski}} \affiliation{\Grenoble}
\author{V. \surname{Kolkovsky}} \affiliation{\Warsaw}
\author{G. \surname{Karczewski}} \affiliation{\Warsaw}
\author{T. \surname{Wojtowicz}} \affiliation{\Warsaw}

\date{\today}

\begin{abstract}
Polarization-resolved magneto-luminescence, together with simultaneous magneto-transport measurements, have been
performed on a two-dimensional electron gas (2DEG) confined in CdTe quantum well in order to determine the
spin-splitting of fully occupied electronic Landau levels, as a function of the magnetic field (arbitrary Landau level
filling factors) and temperature. The spin splitting, extracted from the energy separation of the $ \sigma^+$ and
$\sigma^-$ transitions, is composed of the ordinary Zeeman term and a many-body contribution which is shown to be
driven by the spin-polarization of the 2DEG. It is argued that both these contributions result in a simple, rigid shift
of Landau level ladders with opposite spins.

\end{abstract}


\maketitle A number of experiments on two-dimensional electron
gases (2DEGs)~\cite{Nicholas88,Usher90,Leadley98,Maude98} clearly
show that the thermal activation of carriers across the Fermi
energy, located between the spin split Landau levels at odd
integer filling factors ($\nu$), is governed by a gap which can
significantly surpass the single particle Zeeman energy included
in band structure models. This phenomenon, referred to in the
literature as $g$ factor or spin gap
enhancement,~\cite{Fang68,Janak69,Ando74} is thought to be driven
by the spin polarization of a 2DEG and is a primary manifestation
of the interactions between two-dimensional electrons in the
integer quantum Hall effect (QHE) regime. It is a result of the
specific character of the spin-excitation spectra of a 2DEG at odd
integer $\nu$-QHE states~\cite{Bychkov81,Kallin84}. It can be seen
as arising from the contribution of Coulomb interactions
(including exchange terms) to the energy which is required to
remove, or inject, an electron from, or to, a given spin resolved
Landau level (LL).

To date, the effect of the spin-gap enhancement has been generally
limited to
experiments~\cite{Nicholas88,Usher90,Maude98,Dolgopolov97,Wang92,Wiegers97}
which probe the spin splitting at the Fermi level, for QHE states
at exactly odd filling factors. This limitation has been thought
to be overcome with spectroscopic methods such as, for example,
interband optics~\cite{Kukushkin93,Kukushkin96,Potemski98} or
tunneling experiments~\cite{Dial07}, which, within their trivial
description, permit to investigate the processes of
removing/adding an electron from/to a 2DEG, at arbitrary energy,
filling factor and temperature. Among the different spectroscopic
methods, magneto-luminescence measurements has been widely invoked
to investigate electron-electron correlation in the QHE regime,
however, measurements to probe the spin-gap enhancement are rather
scarce~\cite{Kukushkin93,Dial07}.

Here, we report on magneto-photoluminescence studies of a 2DEG
confined in a high quality CdTe quantum well, and, show that the
enhancement of the spin splitting is not only a property of spin
excitations at the Fermi level, but that it is also relevant for
fully occupied spin Landau levels, located well below the Fermi
energy. We have measured the many body contribution to the spin
gap for fully populated spin Landau levels over a wide range of
filling factors and temperatures, and show that it is driven by
Coulomb interaction, apparent via the spin polarization of the
investigated 2DEG with its relatively large bare Zeeman splitting.

\begin{figure}[h]
\vspace*{-0.5cm}
\begin{center}
\includegraphics[width= 8cm]{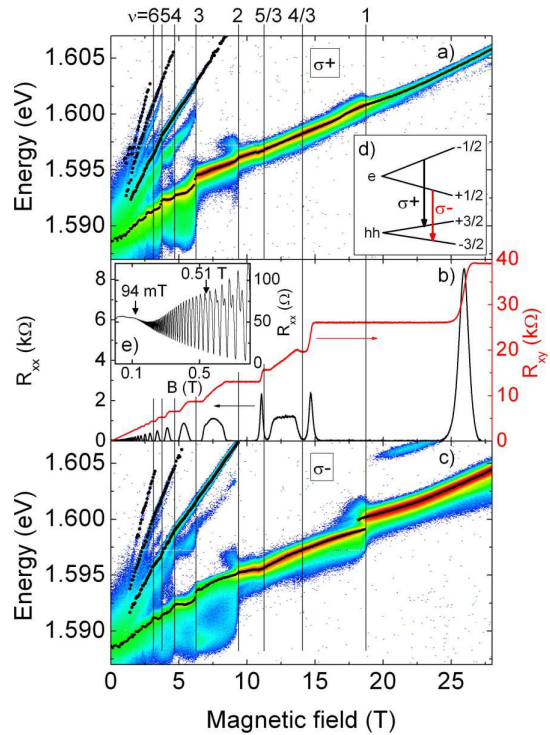}
\vspace*{-0.5cm} \caption{(a) and (c) Color plot of $\sigma^+$ and $\sigma^-$ magneto-PL of a 2DEG in a CdTe QW,
measured at 80~mK, under low power ($\approx0.5$W/m$^2$), $\lambda=514$~nm-Ar$^{+}$ excitation. Black points indicate
the energy of the main peaks. Inset (d) shows the optical selection rules. (b) Results of simultaneous
magneto-transport measurements showing the longitudinal ($R_{xx}$) and Hall ($R_{xy}$) resistance. Vertical lines
indicate the Landau level filling factor. Inset (e) shows an expanded view of $R_{xx}$ at low magnetic fields.}
\label{Fig1}
\end{center}
\vspace*{-1cm}
\end{figure}

The increasingly high quality of GaAs/GaAlAs structures has been
driving advances in the physics of interacting 2D electrons.
Notably, 2D electrons in a GaAs matrix are characterized by a
relatively small bare $g$ factor (-0.44) and therefore by a small
value of the interaction parameter $\eta=E_z/\mathcal{D}$, where
$E_z=g\mu_BB$, $\mathcal{D}=e^2/ \epsilon l_B$, and,
$l_B=\sqrt{\hbar/eB}$ is the magnetic length. The small value of
$\eta$ is responsible for the rich physics exhibited by
interacting 2D electrons in the QHE regime, for example the
occurrence of competing spin polarized/unpolarized many body
ground states~\cite{Clark89} or Skyrmion-type spin texture
excitations~\cite{Sondhi93,Schmeller95,Maude96}. However, this
complex physics often masks the appearance of simpler and basic
many body effects, which should emerge more clearly when $\eta$ is
sufficiently large. Disorder is an additional source of
complications in ascertaining the spin polarization in systems
with small $g$ factors. While high electron mobilities are
obviously advantageous, GaAs-based structures are also rather
fragile, displaying, for example, metastable effects upon
illumination, with an associated decrease in mobility and
homogeneity, which frequently prevents the simultaneous basic
characterization of such structures using magneto-optics and
magneto-transport. A 2DEG in a CdTe matrix~\cite{Karczewski98},
used in our experiments, is characterized by relatively large
(bare) $g$ factor (-1.6) and the $\eta$-parameter in this system
exceeds by a factor of $\approx 3$ its value in GaAs structures
(the dielectric screening $\epsilon=10$ is slightly less efficient
in CdTe). CdTe, which has a conduction band as simple as the one
in GaAs, appears to be an almost ideal model system to study the
QHE physics of the primary spin-polarized states. The significant
progress in the crystal growth of CdTe quantum wells permits
nowadays to attain a 2DEG with reasonably high mobilities. As
shown in Fig.~\ref{Fig1}, the sample studied here shows a well
pronounced fractional QHE and permits a trouble-free, simultaneous
measurement of high quality magneto-photoluminescence and
magneto-transport.

The active part of the investigated structure consist of a
$20$~nm-wide CdTe quantum well (QW), modulation doped on one side
with iodine, and embedded between Cd$_{0.74}$Mg$_{0.26}$Te
barriers. The sample, in form of 1.5$\times$6mm rectangle, was
equipped with electrical contacts in a Hall bar configuration to
permit simultaneous optical and electric measurements. Experiments
have been carried out using either a $^3$He/$^4$He dilution
refrigerator or a variable temperature $^4$He cryostat, in
magnetic fields supplied by a resistive (28~T) or superconducting
(11~T) magnets. A standard, low frequency ($\approx 10$~Hz)
lock-in technique has been applied for the resistance
measurements. Polarization resolved, $\sigma^{+}$ and $\sigma^{-}$
photoluminescence (PL) spectra have been measured using a single
600~$\mu$m-diameter optical fiber to transmit the excitation beam
(514~nm-line of Ar$^+$ laser) and to collect the photoluminescence
signal for the spectrometer (spectral resolution $\approx100\mu
eV$) equipped with a CCD camera. An appropriate linear polarizer
and $\lambda$/4-plate were placed directly between the end of the
fiber and the sample. The $\sigma^{+}$ and $\sigma^{-}$ PL
components were measured by reversing the polarity of the magnetic
field. Special attention has been paid to assure a low level of
laser excitation ($\approx50$~$\mu$W/cm$^2$), to precisely
calibrate the magnetic field, and to measure the spectra at small
intervals (down to 5~mT) of the magnetic field. Under our
experimental conditions (continuous laser illumination), the 2DEG
density of $\approx4.5\times 10^{11}$~cm$^{-2}$ and mobility of
$\mu=2.6 \times 10^{5}$~cm$^2$/Vs were well reproduced in
different experimental runs.

The representative results of simultaneous magneto-PL and
magneto-resistivity measurements of our sample are shown in
Fig.~\ref{Fig1}. As can be seen in Fig.~\ref{Fig1}(b), the
investigated 2DEG shows all typical attributes of the QHE in a
system with fairly high mobility and relatively high electron
concentration; well developed integer QHE states and the
appearance of $5/3$, $4/3$ and $2/3$ fractional states (which will
be discussed elsewhere). From the field at which the Shubnikov de
Haas (SdH) oscillations ($B_{1}\approx94$~mT), and spin-splitting
appears ($B_{2}\approx0.51$~T), we obtain a first estimate of the
enhanced $g$ factor, $g^*\approx3.7$ using the condition ($\hbar
eB_1/m^{*} \approx g^{*}\mu _{B}B_2$) where the electron effective
mass $m^{*}=0.1m_{e}$ was derived from cyclotron resonance
absorption measured on a parent sample. A Dingle analysis of the
SdH oscillations gives a quantum lifetime
$\tau_q=\hbar/2\Gamma=(3.0\pm0.3)$~ps (broadening of Lorentzian
Landau levels $\Gamma\approx110~\mu$eV) as compared to the
transport lifetime $\tau_{\tau}\approx15$~ps (derived from the
measured mobility).

The evolution of the PL with the magnetic field (Fig.~\ref{Fig1})
resembles spectra reported in numerous PL investigations, widely
applied in the past to GaAs-based structures~\cite{Asano02}. Peaks
in the magneto-PL spectra are due to the recombination of
electrons from occupied conduction band LLs ($L_{N},
E_{N}=(N+1/2)\hbar\omega_{c}, N=0,1,..$) with photo-excited holes
from valence band LLs ($L^{h}_{N},
E^{h}_{N}=(N+1/2)\hbar\omega^{h}_{c}, N=0,1,...)$, where
$\omega_{c}$ and $\omega^{h}_{c}$ is the cyclotron frequency of
the electrons and holes respectively. The energy of the main
peaks, due to $L_{N} \rightarrow L^{h}_{N}$ ($N_{e}-N_{h}=0$)
transitions which scale as
$E_{0}+(N+1/2)(\hbar\omega_{c}+\hbar\omega^{h}_{c})$, are shown as
black dots in Fig.~\ref{Fig1}. Since
$\hbar\omega^{h}_{c}/\hbar\omega_{c}\approx5$~\cite{Romestain80}
the magneto-PL spectra reflect largely the characteristic fan
chart of electronic LLs (with respect to band-edge energy,
$E_{0}$) including their occupation factor. Opposite LL spin
components are resolved in the $\sigma^{+}$ and $\sigma^{-}$
spectra. The exchange of the intensity between the $\sigma^{+}$
and $\sigma^{-}$ PL when sweeping through  filling factor $\nu=1$
is typical of the 2DEG studied here and results from  the
selection rules which are specific to CdTe (see
Fig.~\ref{Fig1}(d)).

The non-monotonic variation, with magnetic field, of the
transition energies and intensities (oscillations which correlate
with filling factor) and possible appearance of line splitting
(see Fig.~\ref{Fig1}) are other common features of magneto-PL
investigations of a 2DEG. Electron-electron interactions, combined
with different perturbations induced by the presence of the
valence band hole, are almost certainly at the origin of these
features~\cite{Asano02}. The understanding these features is far
from universal and a detailed analysis of the energy and intensity
of each individual magneto-PL transitions is beyond the scope of
our paper. We have found, however, that information on the effects
of electron-electron interactions can be extracted from the
relative positions of polarization-resolved PL peaks arising from
different LL spin components.

\begin{figure}[t]
\vspace*{-0.5cm}
\begin{center}
\includegraphics[width= 8cm]{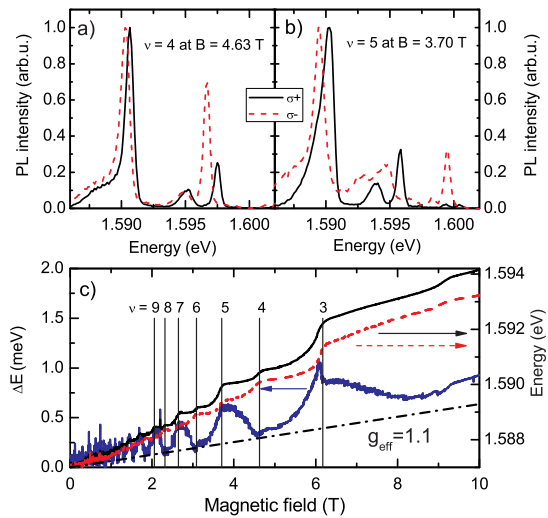}
\vspace*{-0.5cm} \caption {(color online) Normalized $\sigma^+$
and $\sigma^-$ PL spectra at magnetic fields corresponding to
filling factors $\nu=4$ (a) and $\nu=5$ (b) ($T=80$~mK,
$E_{exc}=2.41$~eV, $P_{exc}=0.5$~W/m$^2$). c) Magnetic field
dependence of the energy of polarization resolved transitions from
the different spin levels of the $N=0$ electronic LL (right scale)
and their splitting (left scale). Dashed dotted line shows the
ordinary Zeeman effect in the conduction and valence bands.}
\label{Fig2}
\end{center}
\vspace*{-1cm}
\end{figure}

We focus our attention on the two lowest energy $\sigma^{+}$ and
$\sigma^{-}$ magneto-PL transitions (Fig.~\ref{Fig2}) which are
due to electrons, with different spins, recombining from the fully
populated ($L_{0}$) LL. While the energy of each of these peaks
displays a non-trivial dependence on the magnetic field, here we
focus on the evolution of the energy separation $\Delta E$ between
the $\sigma ^{+}$ and $\sigma ^{-}$ transitions plotted in
Fig.~\ref{Fig2}(c). The splitting $\Delta E$ does not follow a
linear field dependence which is expected for the case of an
ordinary Zeeman effect. This can be even seen in the raw data in
Fig.~\ref{Fig2}(a-b); The splitting $\Delta E$ observed at higher
field $B=4.63$~T is clearly smaller than the splitting at lower
fields $B=3.7$~T. The $\Delta E$ versus $B$ dependence in
Fig.~\ref{Fig2}(c) naturally suggests that this dependence is
composed of two terms; a Zeeman term ($\Delta E _{Z}$) linear with
$B$ and a many body term ($\Delta E _{\uparrow\downarrow}$), which
is non-monotonic with $B$, having maxima at odd integer $\nu$ and
zeros at even integer $\nu$. The linear term can be extracted from
the splitting at even integer $\nu$ and it is in agreement with
the ordinary Zeeman effect expected in our structure. Taking into
account the selection rules depicted in Fig.~\ref{Fig1}(d), the
splitting $\Delta E _{Z}=(|g|-|g_{h}|)\mu_{B}B=g_{eff}\mu_{B}B$,
which requires $g_{eff}=1.1$ to fit the data in agreement with the
reported values of $g=-1.6$ and $g_{h}\approx0.5$, for electronic
and valence hole $g$ factors in CdTe QWs~\cite{Zhao96}.

\begin{figure}[t]
\vspace*{-0.0cm}
\begin{center}
\includegraphics[width= 6cm]{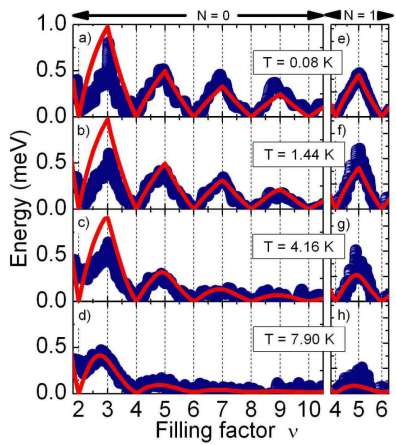}
\end{center}
\vspace*{-0.5cm} \caption{(color online) Many body contribution to
the spin gap in the $N=0$ (left panel) and $N=1$ (right panel)
Landau level extracted from magneto-PL experiments (solid black
points) at different temperatures. Traces calculated using the
model described in the text (red solid lines) are shown for
comparison.}
 \label{Fig3}
\vspace*{-0.5cm}
\end{figure}

To further clarify the origin of the $\Delta
E_{\uparrow\downarrow}$ term, we plot this term as a function of
filling factor and show its characteristic evolution with
temperature (Fig.~\ref{Fig3}). We have extracted $\Delta
E_{\uparrow\downarrow}$ from different experimental runs, by
subtracting the ordinary Zeeman term which is assumed to be
temperature independent. The electron concentration (filling
factor scale) was determined using simultaneous magneto-resistance
measurements. An inspection of the results presented in
Fig.~\ref{Fig3} strongly suggest that $\Delta
E_{\uparrow\downarrow}$ is ruled by the spin polarization
$\mathcal{P}=\frac{n_\downarrow-n_\uparrow}{n_\downarrow+n_\uparrow}$
of the 2DEG. A quantitative verification of this hypothesis is
provided by the following simple model. We consider the ideal case
of a 2DEG with discrete Landau levels separated by
$\hbar\omega_{c}$ and spin split by
\begin{equation}\label{Spingap}
\Delta_{s}=|g|\mu_{B}B + \Delta
E_{\uparrow\downarrow}=|g|\mu_{B}B+
\Delta_{0}'\varphi(B)\cdot\frac{n_\downarrow-n_\uparrow}{n_\downarrow+n_\uparrow}.
\end{equation}
In particular, we assume that the enhanced part ($\Delta E
_{\uparrow\downarrow}$) of the spin splitting is common for all
Landau levels, including the lowest LL ($L_{0}$) which we probe
with PL and the LL in the vicinity of the Fermi energy, the
occupation of which determines the spin polarization. Furthermore,
we suppose that $\varphi(B)=\sqrt[4]{B^{2}+B^{2} _{0}}$ in order
to phenomenologically account for the expected behavior of $\Delta
E_{\uparrow\downarrow}$ in the limit of high magnetic fields
($\varphi(B) \sim \sqrt{B}$) and when $B$ tends to zero
(($\varphi(B)$=constant)~\cite{NOTE}. Finally, we
self-consistently calculate $\Delta E _{\uparrow\downarrow}$ (and
$\mathcal{P}$) and obtain agreement with the data by adjusting the
two fitting parameters, $\Delta_0=\Delta_0'\sqrt{B_{0}}=2.1$~meV
and $B_0=3.7$~T.

Despite the rather crude approximations, the calculations well reproduce the experimental data (Fig.~\ref{Fig3}) over a
wide range of filling factors ($4\leq \nu \leq 10$) and for different temperatures up to the temperature for which
$E_{\uparrow\downarrow}$ (and $\mathcal{P}$) vanishes. The agreement is less satisfactory in the vicinity of $\nu=3$,
and completely fails around $\nu=1$ where difference between $\sigma^+$ and $\sigma^-$ peaks shows almost no
enhancement effect. These discrepancies are due to the fact that the physics of PL processes for a 2DEG at low filling
factors is far more complex~\cite{Asano02} compared to our temptingly simple picture of electrons which recombine (are
extracted) from the homogenous Fermi sea of a 2DEG.

The use of discrete LLs in our calculations is justified by the
large bare Zeeman energy which exceeds the LL width (110~$\mu$eV,
extracted from low field transport data) already at fields of
$\sim2$~T ($\nu \sim 9$). The assumption that
$E_{\uparrow\downarrow}$ does not depend on LL index is probably
also realistic. When modelling the data, we have investigated
various scenarios for a LL index dependence of the spin-gap
enhancement but found that a constant value reproduces the data
fairly well. Although it is more difficult to accurately determine
the energy of the weak magneto-PL peaks for the higher $N>0$ LLs,
it is possible to follow the separation between $\sigma+$ and
$\sigma-$ transitions associated with the $L_{1}$-level in the
vicinity of $\nu=5$. As shown in Fig.~\ref{Fig3} (right panel),
the extracted enhancement of the spin gap in the $L_{1}$-level is
practically the same as in the $L_{1}$-level. Moreover, we find a
fair agreement between the spin gaps $|g|\mu_{B}B + \Delta
E_{\uparrow\downarrow}$, extracted from PL, for fully populated
LLs and the activation gaps, of $0.95$, $0.63$ and $0.36$~meV, for
spin excitations across the Fermi energy, which we have estimated
from resistance measurements at filling factors $\nu=5,7$, and
$9$, respectively.

Finally, let us speculate about a possible extension of the
assumed model to the limit of low magnetic fields and to the
particular case of $\nu=1$. When $B\rightarrow0$, the
extrapolation of Eq.~\ref{Spingap} (at $T=0$~K) yields a linear
$\Delta_{S}$ versus $B$ dependence;
$\Delta_{S}=|g|\mu_{B}B+\Delta_{0}/\nu=(|g|+\Delta_{0}/\mu_{B}B_{\nu=1})
\mu_{B}B =g^{*}\mu_{B}B$, where $B_{\nu=1}$ corresponds to the
magnetic field for $\nu=1$. With $B_{\nu=1}=18.5$~T ($n=4.5\times
10^{11}$~cm$^{-2}$) and $\Delta_{0}=2.1$~meV we extract
$g^{*}=3.6$ for the enhanced $g$ factor in good agreement with the
estimation of $g^{*}\sim3.7$ from the low field onset of spin
splitting in the SdH oscillations. Setting $\nu=1$ (and
$\mathcal{P}=1$) in Eq.~(~\ref{Spingap}) we extrapolate
$\Delta_{S}=|g|\mu_{B}B+\Delta_{0}\sqrt[4]{1+B^{2} _{\nu=1}/ B^{2}
_{0}}$ and calculate $\Delta_{S}=6.4$~meV. This value is a factor
of $\sim4$ smaller than its ultimate limit of
$\sqrt{\pi/2}e^2/\epsilon l_B$~\cite{Bychkov81,Kallin84} but in
good agreement with the reported values in GaAs structures from
optical and capacitance measurements
~\cite{Kukushkin96,Dolgopolov97}.

In conclusion, spectroscopic polarization-resolved magneto-PL
studies of a 2DEG confined in CdTe quantum well reveal the
many-body enhancement of the spin-splitting of fully occupied 2D
Landau levels well below the Fermi energy. The enhancement is
mainly determined by the spin polarization of the 2DEG, since the
spin gap is maximized at odd filling factors, but vanishes at even
filling factors or high temperatures. We argue that the spin
polarization simply induces, in addition to the ordinary Zeeman
splitting, a rigid shift of the spin up Landau levels with respect
to spin-down Landau levels. This simple picture for the many body
spin-gap enhancement emerges from magneto-PL studies of a 2DEG
with relatively large (single particle) $g$ factor.

\acknowledgements{We thank M. Orlita C. Faugeras and R. Grill for
helpful discussions, and acknowledge the financial support from
EC-EuroMagNetII-228043, EC-ITEM MTKD-CT-2005-029671,
CNRS-PICS-4340, MNiSW-N20205432/1198, ERDF-POIG.01.01.02-00-008/08
and SVV-2010-261306 grants. One of us (P.P.) is financially
supported by the EU under FP7, contract no. 221249 `SESAM'.}

\end{document}